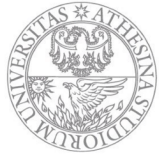

**UNIVERSITY OF TRENTO - Italy**
Department of Psychology
and Cognitive Science

**Master's Degree in Cognitive Science**

**Functional Connectivity in Default Mode Network During Resting State: An Evaluation of the Effects of Data Pre-processing**

*Tutors* *Student*
*Dr. Jorge Jovicich* *Pouya Ghaemmaghami*
*Dr. Domenico Zacà*

Academic Year 2012/2013

# Abstract

# Functional Connectivity in Default Mode Network During Resting State: An Evaluation of the Effects of Data Pre-processing

Resting state functional connectivity estimates from MRI measures has become a promising tool to characterize human brain networks. There are, however, limitations in the method since several sources of errors have been seen to significantly affect the final estimates. This has lead to a great interest in the field to do systematic investigations that help determine the most reliable and robust strategies to perform functional connectivity analysis.

In the present study, we examine the influence of two aspects of data pre-processing in resting state functional connectivity analysis: the effect of criteria used to select nodes in the default mode network (DMN) for the computation of connectivity, and the effect of using or not physiological noise correction. Three different strategies of region of interest (ROI) selection were compared to define DMN node coordinates: (1) ROIs centered on atlas-based coordinates, (2) ROIs based on the result of group independent component analysis, and (3) ROIs based on the estimated DMN of each individual. The study was done using data of 15 healthy volunteers, which had resting state data available from a separate project.

We found that both effects, ROI selection criteria for DMN nodes and physiological noise correction, have significant effects on the functional connectivity estimates. In particular, our results show that physiological noise correction introduces small but significant reductions in functional connectivity, consistent with a reduction of artifactual non-neural correlations introduced by physiological effects. Further, selecting DMN nodes based on the single subject ICA results introduced small but significant increases in functional connectivity, consistent with higher subject specificity of the node selection.

Overall these results contribute to stress the importance of the choice of pre-processing details in functional connectivity analysis, which is a key issue for correctly comparing and interpreting results in the literature.

# Table of Contents

# List of Figures

# List of Tables

# INTRODUCTION

The brain is formed by complex interconnected networks. Functional communication between different parts of these networks has an important role in cognitive processes. So studying functional connectivity in brain networks can provide new insights about the organization of the human brain (Rubinov & Sporns, 2010; Sporns, Chialvo, Kaiser, & Hilgetag, 2004; van den Heuvel & Hulshoff Pol, 2010).

Functional connectivity represents the relationship between the activation patterns of brain areas that are anatomically separated. To date, several methods have been proposed to process functional connectivity in resting-state fMRI data, these methods can be categorized into two groups: model-dependent and model-free methods. In model-dependent methods: regions are generally defined a priori based on the activation maps of the task-dependent fMRI experiments. This predefined ROI is called seed. But for investigating the brain functional connectivity patterns, without defining an a priori seed region, we need to use model-free methods. Model-free methods search for general patterns of connectivity across brain without any pre-assumption or a priori defined ROIs. Several model-free methods have been introduced including principal component analysis (PCA), independent component analysis (ICA) and clustering. Among these methods, ICA is the most common one. ICA separates independent sources of a signal by looking for the existence of sources of signals that are maximally independent from each other (Hyvärinen & Oja, 2000; van den Heuvel & Hulshoff Pol, 2010).

The functioning of the human brain during rest (when individuals are left to think to themselves undisturbed) has been investigated using different techniques (Biswal, Yetkin, Haughton, & Hyde, 1995; De Luca, Beckmann, De Stefano, Matthews, & Smith, 2006; Raichle et al., 2001). Signal fluctuations in fMRI images at rest have been shown to be coherent across different parts of the brain (Biswal et al., 1995; Damoiseaux et al., 2006). Regions showing consistent fluctuations constitute a ''resting-state network'' (RSN) (De Luca et al., 2006).

ICA is a powerful tool for analyzing the functional connectivity in resting-state fMRI data and it can find resting-state networks RSNs that are independent from each other. Regions inside the RSNs can be anatomically separated, however they are functionally related.

Overall, functional connectivity is a powerful tool for providing information about brain organization and in particular large-scale brain networks. However, there are some constraints that we should consider. One of the limitations, As Horwitz (Horwitz, 2003) mentioned, is "lack of uniqueness" in defining and measuring functional connectivity by investigators. This problem comes from the fact that the relationship between different ways of defining and measuring functional connectivity and their underlying neural mechanism is not completely known (Horwitz, 2003). Generally, correlation has been the main method for measuring functional connectivity between brain areas. However, this correlation value only shows the statistical relationship between time-series of different brain regions; so the absence of any types of relation between some brain regions does not necessarily mean the absence of any interaction between those regions (Fingelkurts, Fingelkurts, & Kähkönen, 2005).

Another limitation comes from the underlying structure of these functionally connected regions. In particular, it's not clear whether high correlation between activity patterns of anatomically separated regions reflects anatomical circuits between those regions or not. Anyway this technique has the advantage to generate hypotheses about the potential tracts between anatomically separated regions, which are functionally connected. But this should be studied by other methods to resolve ambiguities (Buckner, Krienen, & Yeo, 2013).

## The DMN (Default Mode Network)

One of the resting states networks that have been widely studied is the default mode network (DMN). Default mode network is a set of brain regions that are typically found deactivated during a wide range of cognitive or goal-directed tasks (Raichle et al., 2001). Although most cognitive tasks cause task-induced deactivation within DMN, there are number of tasks that have been shown to cause increased activity within the default mode network. These tasks require mental simulation of alternative perspectives or imagined scenes like autobiographical memory, envisioning the future, theory of mind and moral decision-making (Buckner, Andrews-Hanna, & Schacter, 2008; Greicius, Supekar, Menon, & Dougherty, 2009). Horovitz et al, have also shown that correlations between frontal (MPFC) and posterior areas (PCC) of the DMN will be reduced during deep sleep (Horovitz et al., 2009).

The default mode network had shown to be disrupted in some disease like autism spectrum disorders (ASD), schizophrenia, Alzheimer's disease (AD), Attention deficit hyperactivity disorder (ADHD) and Depression (Buckner et al., 2008; Greicius, Krasnow,

Reiss, & Menon, 2003; Hedden et al., 2009; Sheline, Barch, Price, Rundle, & Vaishnavi, 2009; Uddin et al., 2008).

Core regions of the DMN include the precuneus/posterior cingulate cortex, dorsal and ventral medial prefrontal cortex, lateral temporal cortex, inferior parietal lobule, and parahippocampal cortex (Buckner et al., 2008). These regions are functionally connected to each other. Since most of the studies focused on functional connectivity analysis and BOLD signal correlations among DMN nodes, there are some concerns regarding the underlying structure of these functionally connected area. In this regard, Greicius et al, combined DTI tractography and resting-state fMRI to see whether DMN functional connectivity reflects structural connectivity or not. They showed that there are direct connections between PCC and MTL and between PCC and MPFC while there are no tracts connecting the MPFC to the MTL (Greicius et al., 2009).

## Physiological Noise

Functional Magnetic Resonance Imaging (fMRI) relies on the use of blood-oxygen level dependent (BOLD) contrast to detect brain regions that respond to task-induced activation (Biswal, Yetkin, Haughton, & Hyde, 1995; Chang, Cunningham, & Glover, 2009). However BOLD contrast is an indirect measure of cerebral metabolism since it comes from hemodynamically-driven changes in tissue and vessel oxygenation (Chang et al., 2009). The other sources that also cause changes in cerebral blood flow are physiological processes including blood pulsatility and respiration (Chang et al., 2009).

This artifact reduces the statistical significance of the BOLD signal (Chang et al., 2009; Glover, Li, & Ress, 2000).

Blood pulsatility has some effects on the measured BOLD signal in the areas near to the large vessels, by causing the movement of tissue and also by producing an influx of unsaturated blood into the region of interest (Dagli, Ingeholm, & Haxby, 1999). These cardiac-related artifacts, as Dagli et al has shown, are highly consistent across subjects in magnitude and location and they localize spatially in specific areas such as major blood vessels, sulci, and CSF pools (Dagli et al., 1999; Glover et al., 2000). Since the signal change coming from cardiac-induced noise is equal or greater than the signal change, which is produced by cortical function, the ability to detect neural activity in those specific regions would be decreased (Dagli et al., 1999).

Respiration causes bulk motion of the head and also leads to the modulation of the magnetic field that can cause a shifting of the brain image (Birn, Diamond, Smith, & Bandettini, 2006; Glover et al., 2000). Breathing has another effect on the fMRI time courses. Subtle changes in breathing depth and rate that occur during resting scan can cause $CO_2$ fluctuations, which are correlated significantly with the BOLD fMRI signal fluctuations (Birn et al., 2006; Birn, Smith, Jones, & Bandettini, 2009; Chang et al., 2009; Wise, Ide, Poulin, & Tracey, 2004).

In this regard, several methods have been proposed for reducing the artifacts related to physiological noise in fMRI time-series. One of the most famous methods is RETROICOR (Image-Based Method for Retrospective Correction) proposed by Glover et al., (2000). This method corrects physiological artifacts by fitting low order Fourier series to the image data based on the phase of the respiratory or cardiac cycle during each acquisition (Glover et al., 2000). Figure 1 shows an example of correcting BOLD times-series with RETROICOR.

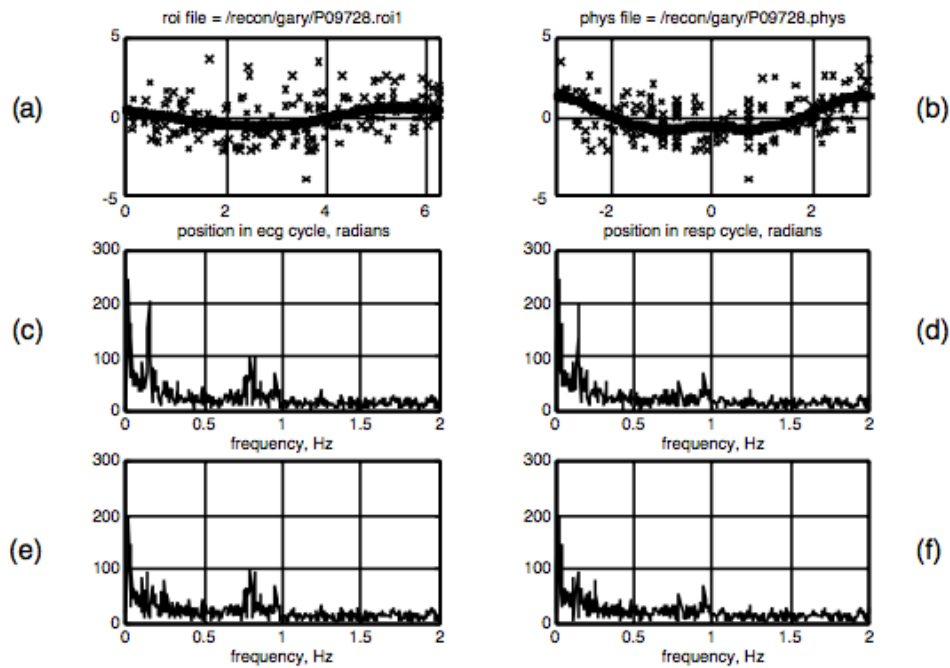

Figure 1: RETROICOR method applied to ROI time-series data acquired at TR = 250ms. Figure 1a shows raw data (+) and cardiac fit (*) plotted versus phase of cardiac cycle; Figure 1b shows the raw data plotted versus phase of respiratory cycle (+) and corresponding respiratory fit (*). Figure 1c shows the spectra of time-series without correction; Figure 1d shows the spectra with cardiac correction; Figure 1e shows the spectra with respiratory correction; Figure 1f shows the spectra with both cardiac and respiratory corrections. (Data from Glover et al., 2000)

In this study, we used a modified version of RETROICOR (8 physiological regressors) with an additional regressor to model the respiratory volume per time as described in Birn et al. (2006).

## ROIs Selection

One of the fundamental and important steps regarding functional connectivity analyses is selecting brain regions, from which signal time courses will be extracted. This procedure is done in different ways. Sometimes its done based on the previous studies, or by choosing areas that show activations or deactivations in response to the particular task.

However, These different ROI selection strategies can lead to different results. Marrelec et al, 2011, investigated the effect of four different ROI selection methods within DMN. They wanted to see that in specific brain regions, how similar are the time courses obtained by different ROI selection approaches, and to which extent changes induced by these different ROI selection methods has an effect on the functional connectivity results. They showed that ROI selection method has an effect on the signal variance while it has no effect on the signal mean. They also showed how changes in the anatomical location of the ROIs make changes in the functional correlation matrix (Marrelec & Fransson, 2011).

For the present study, we focused on the relation between different choices of ROI selection methods on functional connectivity within the DMN during resting state with two different pro-processing approaches.

Three different strategies of ROI selection will be compared: (1) ROIs centered on the coordinates given in (Hedden et al., 2009), (2) ROIs based on group independent component analysis, and (3) ROIs based on performing combination of single subject DMN and ATLAS.

## Aims of the study

As it is shown by Buckner et al., (2013) functional connectivity measures are sensitive to confounding factors (including head motion and physiological artifacts) and because of the importance of the pre-processing in BOLD fMRI time courses, in the present study we want to examine the effects of different pre-processing approaches on resting-state functional connectivity in DMN. In particular we just focused on two steps of the pre-processing pipeline: Physiological correction and ROI selection criteria.

We propose two driving hypotheses: i) physiological noise correction is expected to reduce the contribution of non-neural correlations, thereby reducing functional connectivity, and ii) the increased functional gray matter specificity in the selection of DMN nodes using subject specific ICA should reduce the contribution of unrelated fluctuations, thereby increasing the functional connectivity with other nodes. The aim of this work is to test these two hypotheses, in particular in terms of statistical significance using data acquired at 4T on healthy young volunteers.

# METHODS

## UNITN Resting State study

### SUBJECTS

Participants included 15 healthy adults. One male was excluded because of the excessive movement, so analysis was done based on the remaining 14 subjects (mean age 27.36 ± 4.55 years, 8 males). All subjects provided written, informed consent to participate in this study with no history of psychiatric or neurological diseases. The ethical committee of The University of Trento approved this study.

### DATA ACQUISITION

Images were acquired at 4.0 T using a MRI scanner (Bruker Medical, Ettlingen, Germany) with a birdcage-transmit, 8-channel receive head coil (USA Instruments, Inc., Ohio, USA). EPI BOLD functional images (A total of 215 EPI volumes) were acquired using the point-spread-function distortion correction method (Zaitsev, Hennig, & Speck, 2004) with standard parameters (TR = 2200 ms, TE = 33 ms, interleaved slices, 3×3×3 mm$^3$ isotropic voxel size, matrix size 64×64×37, flip angle = 75°, FOV = 192 cm).

#### RESTING STATE SCAN

All subjects underwent a resting state scan, where they were asked to keep their eyes closed and to let their mind wander freely while trying not to engage into structured thoughts. The duration of the resting state scan was 493 seconds.

#### PHYSIOLOGICAL MONITORING

Cardiac and respiratory processes were monitored during the functional scan using the scanner's built-in physiological recording equipment (photoplethysmograph and respiratory belt). These were both sampled at 50 Hz and saved to a data file.

## Data Preprocessing

Resting state functional data were preprocessed using AFNI (Cox, 1996) and FSL (FSL v5.0.2; FMRIB Oxford University, UK) (Smith et al., 2004; Woolrich et al., 2009).

The first 5 volumes of EPI images were discarded from the analysis to allow signal stabilization due to T1 recovery effect. Then, pre-processing continued using two different approaches.

In the first approach motion correction and slice-timing correction were performed using AFNI 3dvolreg command with –tshift option for slice-timing correction.

The second approach began by just applying motion correction (using 3dvolreg command in AFNI). It was followed by "Physiological Noise" correction on the motion corrected data using the RETROICOR method (Glover et al., 2000) (This will be explained in the next section). Then slice-timing correction was performed on the "Physiological Noise" corrected data (by using 3dtshift command in AFNI).

Registration of the EPI functional images to MNI template was carried out in FSL MELODIC using 12 degrees-of-freedom (full affine). (Jenkinson & Smith, 2001)

High pass filter cutoff was chosen at 0.01 Hz.

**PHYSIOLOGICAL NOISE CORRECTION WITH RETROICOR**

Physiological noise correction consisted of: (1) phase shifting of the physiological data to match the timing of each slice's acquisition using AFNI's retroTS.m procedure; (2) Obtaining13 slice-based regressors based on the recorded cardiac and respiration data. These regressors included 4 regressors regarding to the cardiac series and its harmonics, 4 for the respiratory series and its harmonics and 5 for respiration variation of time and its harmonics (Birn et al., 2006) (3) The variance explained by these 13 regressors was removed from the BOLD time series with RETROICOR procedure implemented in AFNI.

## ROI ANALYSIS

According to the study done by (Hedden et al., 2009), four regions of interest (ROIs) belonging to the default mode network were selected: precuneus/posterior cingulate cortex (PCC), left and right lateral parietal cortices (lLPC, rLPC), medial prefrontal cortex (MPFC). In this study, these four regions were selected in three different ways:

- ➢ Using MNI coordinates based on ATLAS as given in (Hedden et al., 2009) denoted as ATLAS

- Performing group independent component analysis (ICA) denoted as gICA
- Performing combination of subject estimated ICA and ATLAS denoted as Ind-ICA

**ROIs BASED ON ATLAS**

ROIs were defined by creating spherical seed regions with a 12 mm radius centered in four regions: posterior cingulate cortex (PCC) (MNI coordinate: 0, -53, 26) medial prefrontal cortex (MPFC) (MNI coordinate: 0, 52, -6) and two lateral parietal cortices (LPC) (MNI coordinate: -48, -62, 36; 46, -62, 32). These MNI coordinates were obtained from previous studies. (Hedden et al., 2009) ROIs are shown in Figure 2.

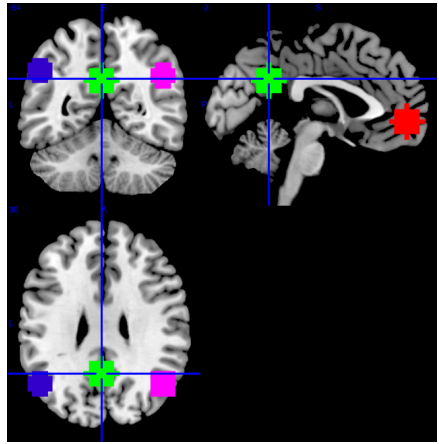

Figure 2: Four ATLAS ROIs corresponding to four main regions of Default Mode Network. Coordinates are in MNI template as given in Hedden et al (2009). The ROI in green is PCC with MNI coordinate: (0, -53, 26). ROI with red color is MPFC with MNI coordinate: (0, 52, -6). ROI with blue color is lLPC with MNI coordinate: (-48, -62, 36). And ROI with pink color is rLPC with MNI coordinate: (46, -62, 32)

**ROIs BASED ON gICA RESULT**

ROI selection based on the result of group independent component analysis was accomplished in a four-step procedure. First, data was variance normalized and the group independent component analysis was performed using the multi-session temporal concatenation approach of the ICA module implemented in FSL MELODIC (Beckmann & Smith, 2004; Beckmann et al., 2005). Second, from the extracted independent components (71 components in the no physiological corrected dataset and 68 components in physiological corrected dataset. The threshold for a voxel to be in the DMN in non physiologically corrected dataset was 2.5 and it was 2.37 in physiologically corrected dataset), the best-fitting component was extracted by matching each independent component with a (DMN) default mode network template as given in (Smith et al., 2009) and (Uddin, Kelly, Biswal, Castellanos, & Milham, 2009) (Figure 5). Third, the Z-score map of the component representing DMN was saved in NIFTI format and it was thresholded to make 4 predefined regions (PCC, MPFC and two LPC) spatially separated and 4 different masks (corresponding to these 4 regions) were created. Forth, in each mask, the voxel with the highest Z-score from the DMN component was determined and a sphere with 12 mm radius centered in the coordinates of that voxel, was created in "mricro (http://www.mccauslandcenter.sc.edu/mricro/)" with two constraints to avoid inclusion of irrelevant areas: maximum image intensity difference between the central seed and other voxels in the ROI was set to be below 20 and likewise maximum brightness difference between voxels in the ROI and the central seed was set to be below 20 (Figure 6a).

**ROIs BASED ON COMBINATION OF SUBJECT ESTIMATED ICA AND ATLAS (IND-ICA)**

ROI selection based on single subject data was accomplished in a three-step procedure. First, for each subject, based on the group ICA result, an estimation of the independent components spatial maps was obtained using a procedure called "Dual Regression" implemented in FSL (Beckmann, Mackay, Filippini, & Smith, 2009; Filippini et al., 2009). This procedure estimates each subject spatial maps in two stages. In the first stage, it uses the group spatial maps and regresses them into each subject's 4D dataset to give a set of time courses. In the second stage, obtained time courses from the first stage are used as a set of temporal regressors to find subject-specific maps for getting subject-specific spatial maps (Figure 3). Second, the Z-score map of the component representing DMN for each subject was saved in NIFTI format and it was thresholded to make 4 predefined regions (PCC, MPFC and two LPC) spatially separated and it was saved as a mask including all 4 regions. Third, four different ROIs were created by intersecting this mask with our four ATLAS based ROIs generated as previously described. Figure 6b shows a sample of final ROIs based on this approach for one of the subjects.

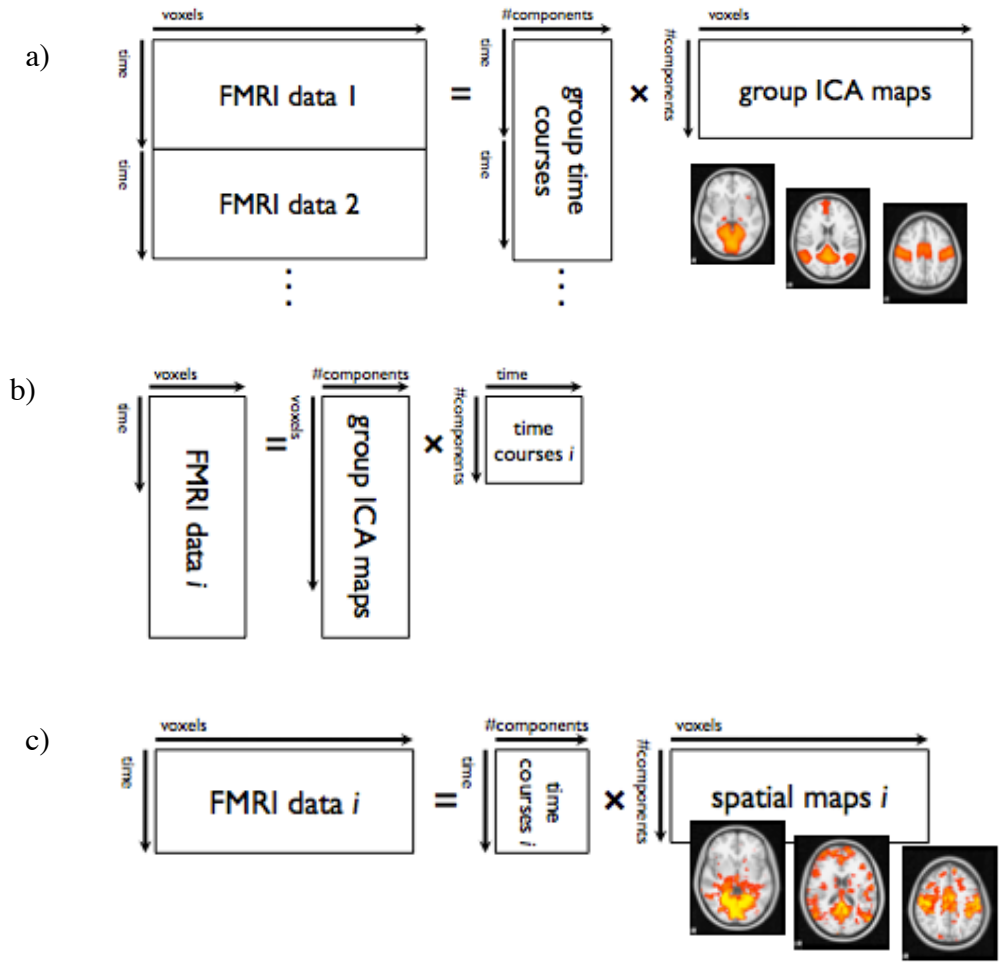

Figure 3: Dual Regression method: (a) FMRI data sets of the whole group of subjects are temporally concatenated and ICA is applied in order to identify large-scale patterns of functional connectivity in the population; (b) group-level spatial maps are used as a set of spatial regressors in a GLM, to find temporal dynamics associated with each group-level map; (c) these time-courses are used as a set of temporal regressors in a GLM, to find subject-specific maps (Data from Beckmann et al., 2009).

## Connectivity Analysis

For analyzing functional connectivity, first, the mean time-series was obtained from all voxels within each of our four network regions (MPFC, PCC, lIPC and rIPC) (Figure 4). Then Pearson's correlation coefficient was calculated between the mean time courses for each pair of ROIs. This was used as a measure of the functional connectivity of the default network.

We obtained 3 (ROI selection methods) × 2 (Pre-processing approaches) × 6 (pairwise correlation between nodes) × 14 (subjects) correlation values to evaluate the variability induced by pre-processing and ROI selection criteria on functional connectivity within DMN. All computations were performed using "R" software (R_Team, 2008).

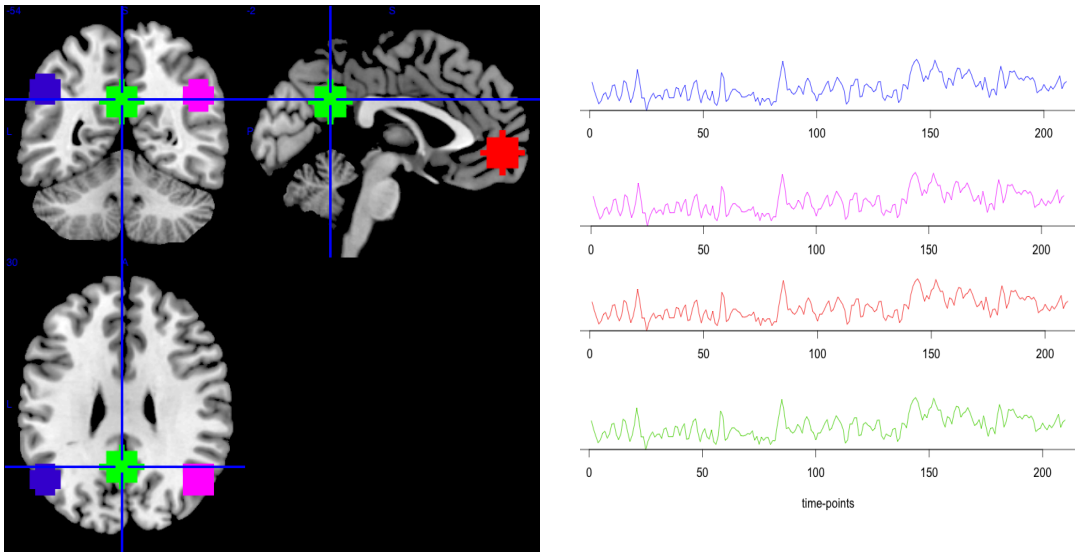

Figure 4: Four DMN regions (MPFC, PCC, lLPC and rLPC) and their corresponding mean time courses from all voxels within each of these four network regions. For analyzing the functional connectivity in DMN, we calculated the pairwise pearson correlation between these ROIs. So correlations between MPFC-PCC, MPFC-rLPC, MPFC-lLPC, PCC-rLPC, PCC-lLPC and lLPC-rLPC were calculated.

# RESULTS

In this study we use resting state fMRI data from 15 healthy subjects to examine 1) the influence of different ROI selection strategies on the variability in anatomical location of the seeds and 2) the effect of pre-processing (physiological noise correction and ROIs selection criteria) on functional connectivity.

## Basic quality assurance of fMRI data

The raw fMRI data was visually inspected for general quality assurance before the analysis. No major visible artifacts were observed (RF interferences, distortions, wrap around).

Recent findings stress the importance of head motion as a source for potential functional connectivity biases, in particular when comparing subject groups who have potential differences in motion degree (Power, Barnes, Snyder, Schlaggar, & Petersen, 2012). Even though in our study we do not do comparisons between subject groups, we inspected the motion correction parameters. We found that, except one subject, for all the other subjects the three head translation and rotation parameters during the resting state fMRI scan were below 1.2 mm and 1 degree. So our analysis has built based on 14 subjects (one subject was discarded because of the excessive movement).

# Effect of ROI selection criteria on the variability in anatomical location of the seeds

Table 1 shows the variability in anatomical location of seeds induced by different ROI selection strategies. Except PCC that its location is consistent among subjects, the center of the gravity of the other regions are different between ROI selection strategies and consequently among subjects.

### MASKS CREATED BASED ON GROUP ICA RESULT

Figure 5a shows the gICA result of DMN for the dataset with no physiological noise correction and Figure 5b shows the result with physiological noise correction. As explained in methods, based on these gICA results, 4 masks corresponding to the 4 regions of DMN (MPFC, PCC, rLPC, lLPC) were created (Figure 6a).

### MASKS CREATED BASED ON IND-ICA APPROACH

Regarding making Ind-ICA mask, as explained in methods, for each subject, an estimation of the independent components spatial maps, based on the group ICA result, was obtained using a procedure called "Dual Regression". Figure 6c shows the DMN estimated component for some of the subjects based on this approach. By intersecting this component (which is saved as a mask) with four ATLAS based ROIs four DMN ROIs were created (Figure 6b).

Table 1 shows the MNI coordinate of the 'center of gravity' of all ROIs: including ATLAS based ROIs, ROIs based on group ICA and ROIs based on Ind-ICA approach.

a) 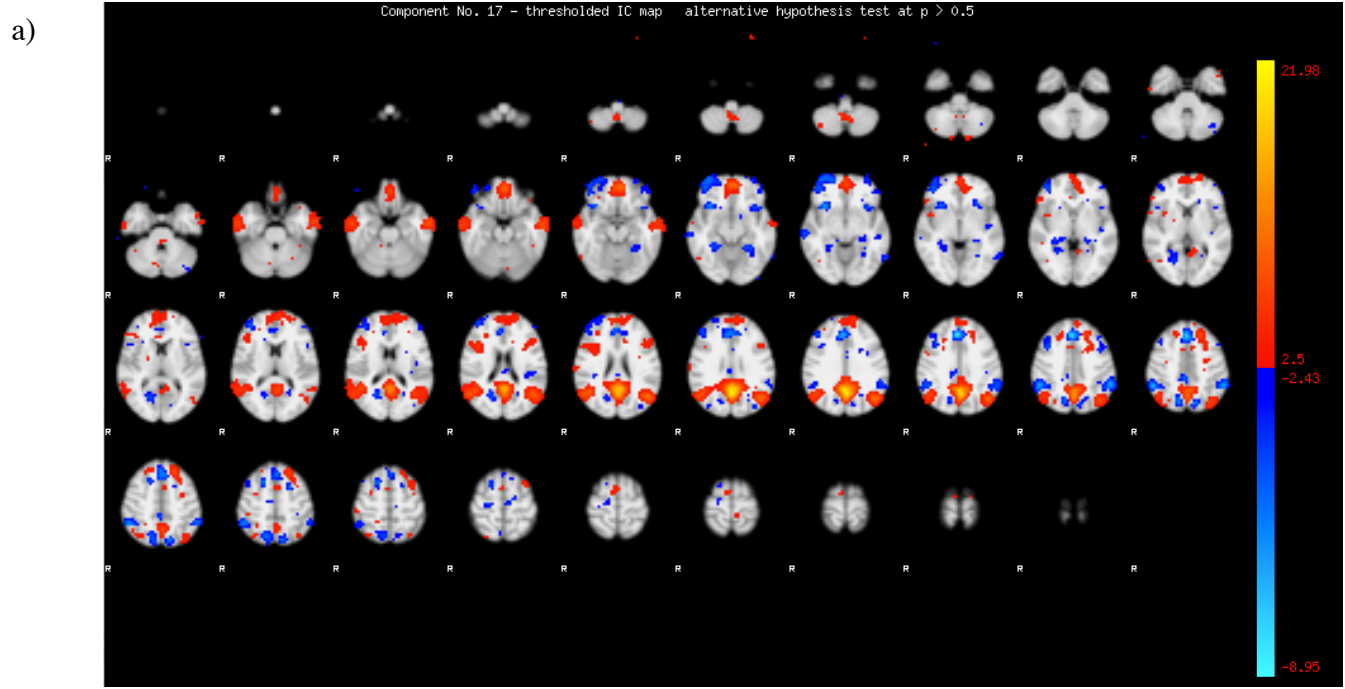

b) 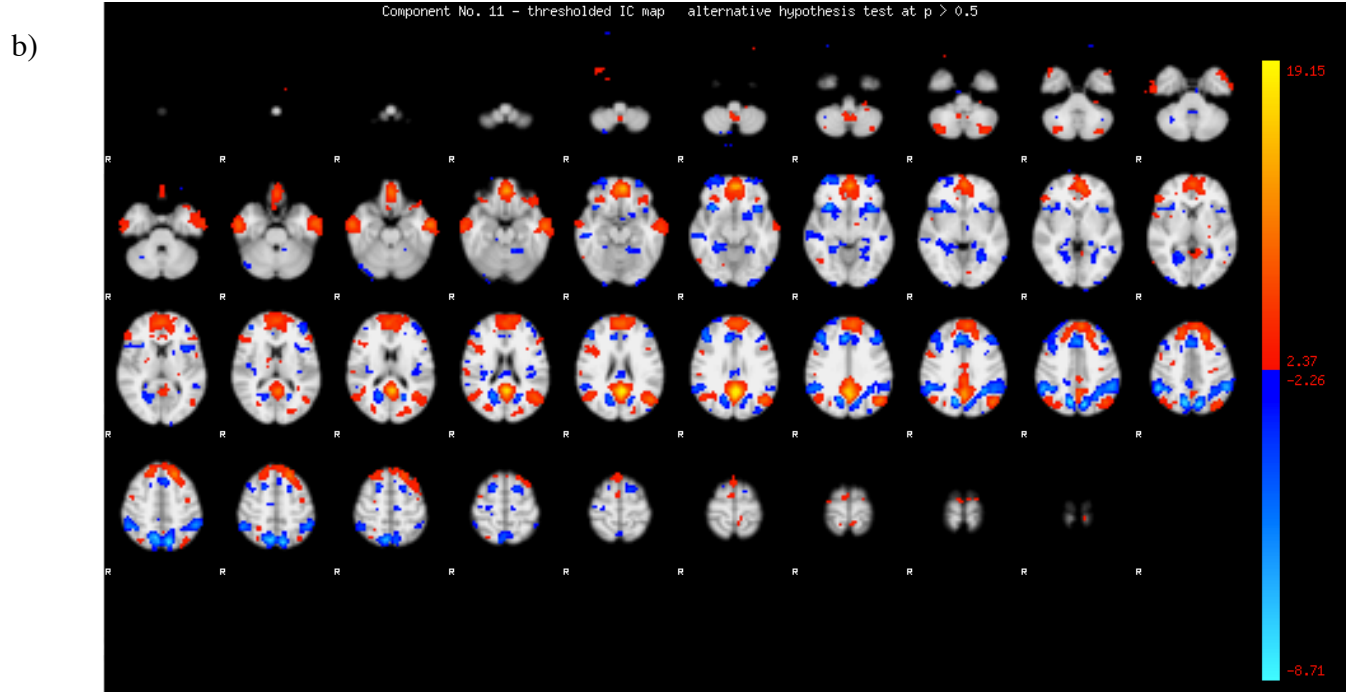

Figure 5: DMN component in gICA. Figure 5a shows DMN component in not physiologically corrected dataset; Figure 5b, shows the DMN component in Physiological Noise corrected dataset.

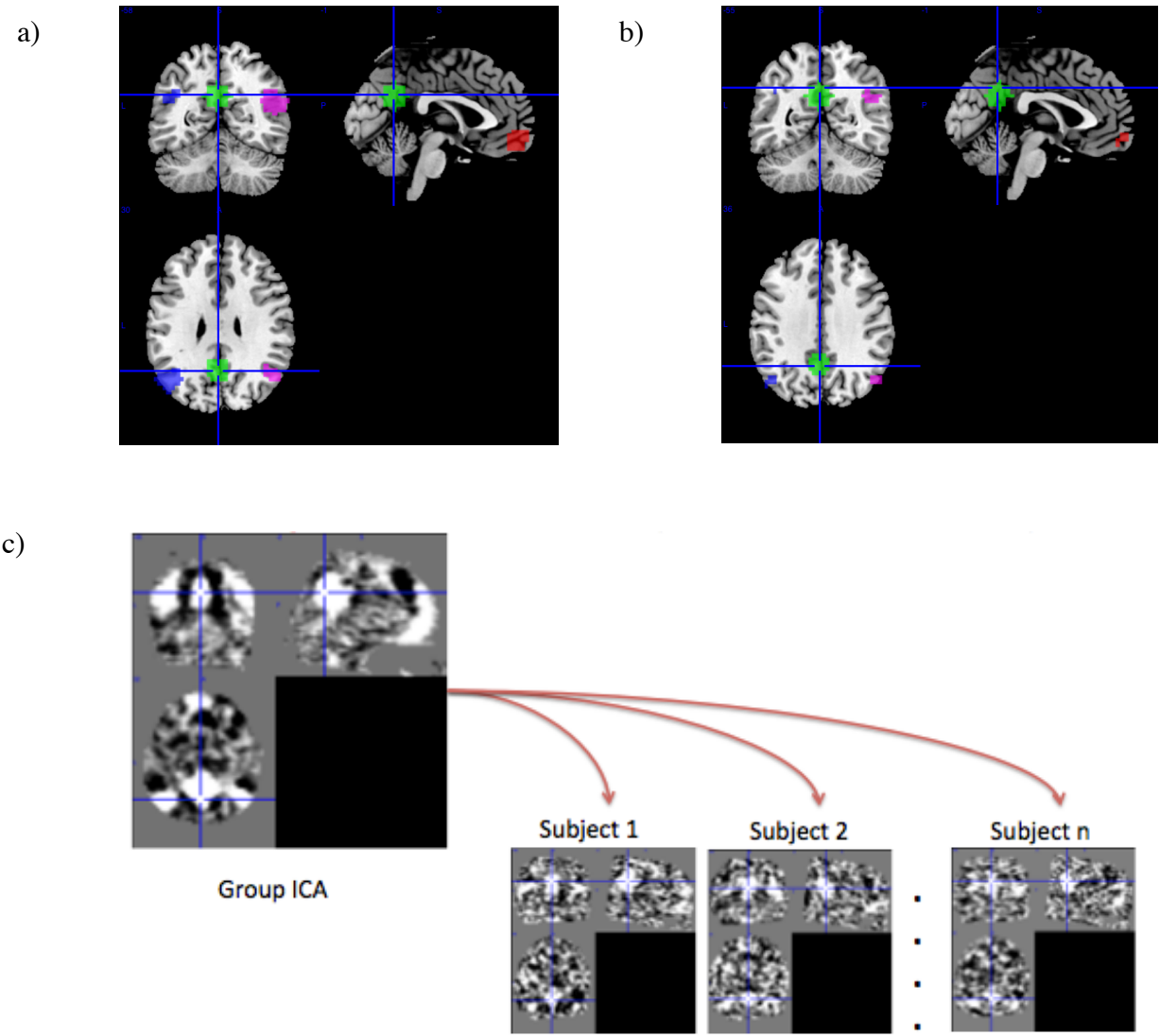

Figure 6: Figure 6a shows the DMN masks created based on the gICA result on the non-physiologically corrected dataset. Figure 6b shows the sample of a DMN masks created based on the Dual regression method for one subject. Figure 3c shows the Dual regression approach.

|         | No Physio |  |  |  | Physio |  |  |  |
|---------|-----------|-----|------|------|--------|-----|------|------|
|         | MPFC | PCC | lLPC | rLPC | MPFC | PCC | lLPC | rLPC |
| **ATLAS** | (-2, 54, -4) | (-2, -54, 28) | (-50, -62, 36) | (46, -62, 32) | | | | |
| **gICA** | (0, 53, -12) | (-2, -58, 28) | (-45, -66, 31) | (50, -59, 23) | (-4, 50, -15) | (-2, -50, 24) | (-47, -67, 31) | (47, -61, 25) |
| **Sub 1** | (-2, 57, -10) | (-2, -54, 28) | (-48, -65, 31) | (48, -62, 31) | (-2, 57, -5) | (-2, -54, 28) | (-49, -66, 31) | (47, -60, 30) |
| **Sub 2** | (-5, 55, -9) | (-2, -54, 28) | (-47, -65, 34) | (47, -59, 31) | (-5, 56, -7) | (-2, -54, 27) | (-48, -66, 35) | (49, -58, 32) |
| **Sub 3** | (-2, 54, -6) | (-2, -54, 28) | (-46, -64, 36) | (48, -63, 30) | (-2, 54, -6) | (-2, -54, 28) | (-47, -66, 35) | (49, -66, 30) |
| **Sub 4** | (-1, 55, -6) | (-2, -54, 28) | (-50, -61, 34) | (49, -61, 33) | (-2, 54, -5) | (-2, -54, 28) | (-52, -62, 33) | (50, -62, 34) |
| **Sub 5** | (-4, 50, -6) | (-2, -54, 28) | (-47, -66, 33) | (43, -61, 32) | (-3, 54, -5) | (-1, -54, 27) | (-48, -66, 34) | (43, -62, 34) |
| **Sub 6** | (-3, 55, -5) | (-2, -54, 28) | (-48, -65, 34) | (49, -64, 25) | (-2, 54, -5) | (-2, -54, 29) | (-50, -65, 33) | (51, -63, 25) |
| **Sub 7** | (-2, 55, -5) | (-2, -54, 28) | (-48, -65, 33) | (46, -60, 29) | (-2, 54, -4) | (-3, -52, 28) | (-48, -65, 34) | (46, -53, 29) |
| **Sub 8** | (-2, 55, -7) | (-2, -54, 28) | (-48, -66, 35) | (48, -62, 25) | (0, 54, -6) | (-2, -54, 28) | (-46, -68, 35) | (46, -60, 26) |
| **Sub 9** | (-4, 55, -4) | (-2, -54, 28) | (-46, -65, 35) | (47, -58, 29) | (-2, 53, -4) | (-2, -54, 28) | (-46, -65, 34) | (48, -59, 29) |
| **Sub 10** | (-3, 58, -2) | (-2, -54, 28) | (-50, -64, 33) | (48, -63, 33) | (-2, 56, -3) | (-2, -54, 27) | (-51, -65, 34) | (48, -62, 34) |
| **Sub 11** | (0, 52, -10) | (-2, -54, 27) | (-50, -63, 34) | (43, -64, 30) | (-2, 52, -6) | (-3, -54, 27) | (-48, -65, 34) | (42, -64, 31) |
| **Sub 12** | (-3, 54, -4) | (-2, -54, 28) | (-47, -65, 37) | (47, -63, 32) | (-2, 54, -4) | (-2, -54, 28) | (-48, -65, 36) | (46, -65, 34) |
| **Sub 13** | (-3, 58, -4) | (-2, -54, 29) | (-48, -66, 32) | (46, -62, 32) | (-4, 55, -4) | (-2, -53, 28) | (-49, -67, 32) | (47, -63, 30) |
| **Sub 14** | (-1, 55, -6) | (-1, -54, 28) | (-49, -64, 36) | (47, -64, 30) | (-2, 53, -6) | (-2, -53, 28) | (-52, -65, 34) | (47, -66, 31) |

Table 1: Center of gravity of each ROI based on different ROIs selection criteria: Atlas, group ICA, and ROIs based on each individual's estimated ICA spatial maps. Node coordinates are given in MNI atlas space (mm).

# Effect of Physiological Noise and ROIs selection criteria on Functional Connectivity

We examined the effect of different BOLD fMRI pre-processing (different ROI selection criteria and correction of physiological noise) on functional connectivity of the default mode network. Regarding this, the pairwise correlations between extracted mean time courses of four DMN regions were computed (see Methods). These correlations were calculated 6 times for each subject (3 criteria for ROI Selection × 2 pre-processing pipelines, with and without physiological noise correction) for each of the 6 possible pairs of ROIs included in the DMN. Figure 7 and Figure 8 shows these results.

For investigating the effects of ROI selection strategies and Physiological noise on functional connectivity we did a 3-way ANOVA (Table 2). The results did not provide evidence of statistically significant interaction effect (F=0.09, p=0.92) between the ROI criteria selection and the choice of whether or not regressing out the physiological noise from the data.

However, the test for the main effect of Physiological Noise demonstrated a statistically significant Physiological noise correction effect on the Functional Connectivity (F=8.51, p=0.004). Figure 7 shows that performing physiological noise correction decreases the functional connectivity (pairwise correlation between ROIs). This effect is consistent in all of the datasets with different ROI selection criteria. But for further analyzing the effect of physiological noise on functional connectivity, we performed a 2-way ANOVA on three separated datasets corresponding to each of the ROI strategies. Results did not show statistically significant effect of physiological noise on functional connectivity (Table 3-5). This suggests that there is a trend, which is not significant when analysis is done separately on each ROI selection datasets, but it reaches at significance level when all of the factors are together.

Likewise, the test for the main effect of the criteria for ROI selection (F=7.06, p=0.001) showed a statistically significant effect on the Functional Connectivity. Figure 8 shows that selecting ROIs based on each individual spatial maps increases the functional connectivity (pairwise correlation between ROIs). This effect is consistent in both datasets with different pre-processing approaches (physiologically corrected dataset and the dataset with no physiological noise correction). Similarly, for examining the effect of ROI selection strategies on functional connectivity, we performed a 2-way ANOVA on two separated datasets corresponding to each of the two pre-processed approaches. Results show a significant effect of ROI selection criteria on functional connectivity (Table 6 and Table 7).

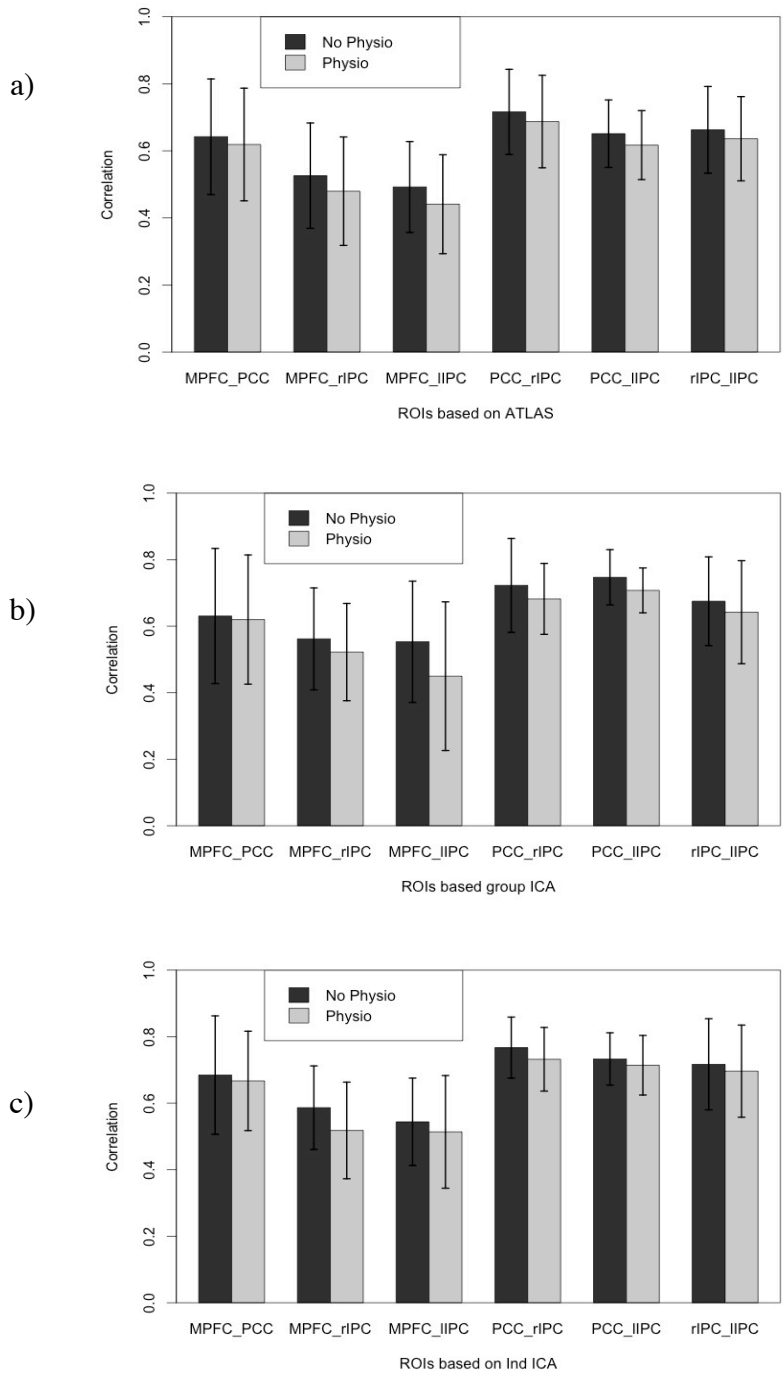

Figure 7: Effect of physiological noise correction on functional connectivity with different ROI Selection strategies. Figure 7a shows this effect where the ROIs were chosen based on ATLAS. Figure 7b shows the effect while regions were chosen based on group ICA. Figure 7c shows this effect while regions were chosen based on individual estimated ICA. On the X-axis the 6 pairs of ROIS included in the DMN are reported. On the Y-axis mean correlation and standard deviation across 14 subjects are reported.

a)

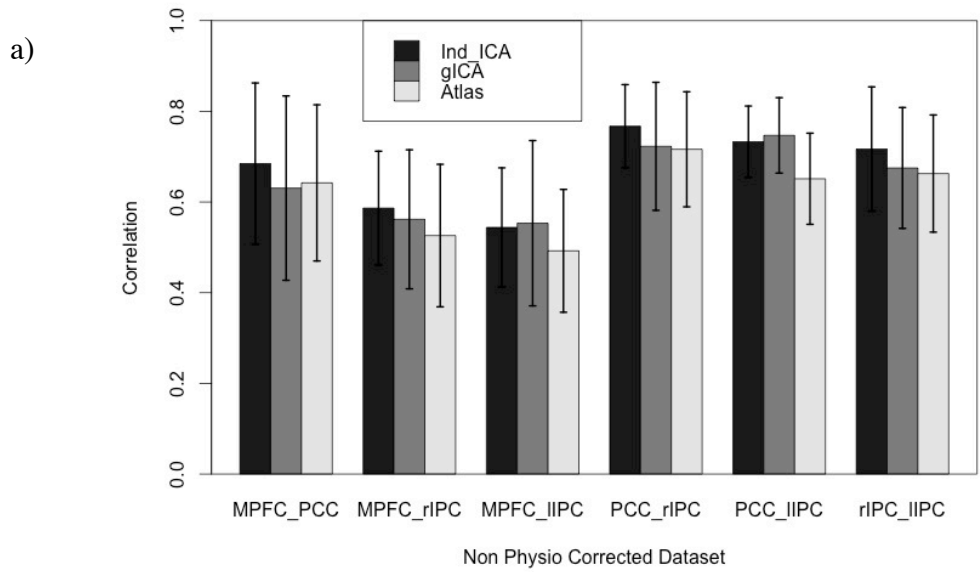

b)

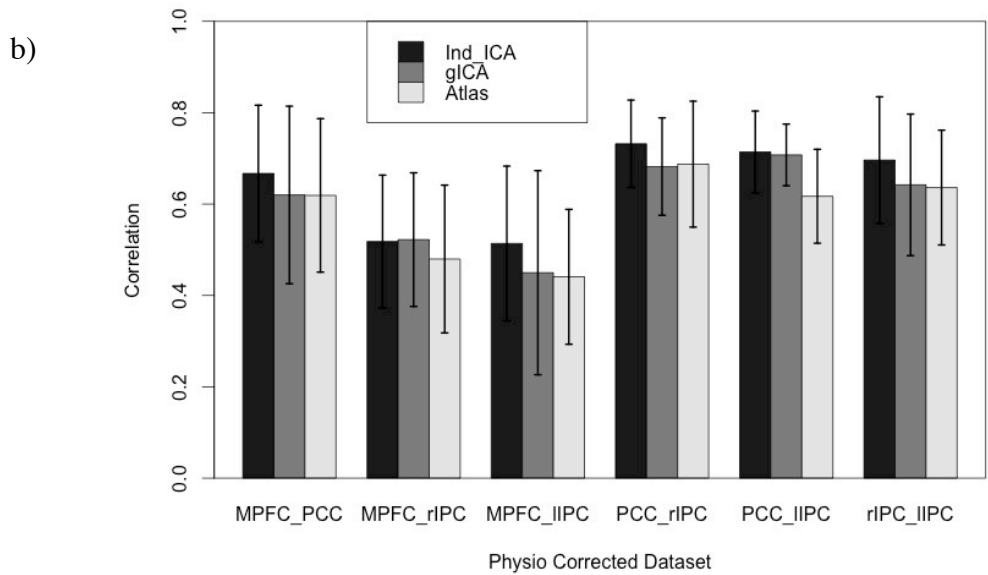

Figure 8: Effect of ROI selection criteria on functional connectivity with two different pre-processing pipelines (with and without physiological noise correction). Figure 8a shows this effect where physiological correction was not done. Figure 8b shows the effect while data was physiologically corrected. On the X-axis the 6 pairs of ROIS included in the DMN are reported. On the Y-axis mean correlation and standard deviation across 14 subjects are reported.

| Effects | DF | Sum Sq | Mean Sq | F value | Pr>F | Significancy |
|---|---|---|---|---|---|---|
| Pre_Processing | 1 | 0.1737 | 0.17371 | 8.5076 | 0.0037067 | ** |
| ROI_Strategy | 2 | 0.2882 | 0.14408 | 7.0563 | 0.0009567 | *** |
| Area | 5 | 3.4024 | 0.68049 | 33.3269 | < 2.2e-16 | *** |
| Pre-Processing:ROI-Strategy | 2 | 0.0036 | 0.00180 | 0.0881 | 0.9157142 | |
| Pre-Processing:Area | 5 | 0.0287 | 0.00574 | 0.2809 | 0.9235703 | |
| ROI-Strategy:Area | 10 | 0.0983 | 0.00983 | 0.4816 | 0.9021128 | |
| Pre-Processing:ROI-Strategy:Area | 10 | 0.0226 | 0.00226 | 0.1106 | 0.9997185 | |
| Residuals | 468 | 9.5559 | 0.02042 | | | |
| | | | | | | |
| Signif. codes | 0 "***" | 0.001 "**" | 0.01 "*" | 0.05 "." | 0.01 " " | 1 |

Table 2: ANOVA results for analyzing the effects of Physiological noise correction and ROI strategy on functional connectivity.

| Effects | DF | Sum Sq | Mean Sq | F value | Pr>F | Significancy |
|---|---|---|---|---|---|---|
| Pre-Processing | 1 | 0.05176 | 0.051765 | 2.6205 | 0.1075 | |
| Area | 5 | 1.17959 | 0.235918 | 11.9430 | 8.522e-10 | *** |
| Pre-Processing:Area | 5 | 0.00453 | 0.000907 | 0.0459 | 0.9987 | |
| Residuals | 156 | 3.08157 | 0.019754 | | | |
| | | | | | | |
| Signif. codes | 0 "***" | 0.001 "**" | 0.01 "*" | 0.05 "." | 0.01 " " | 1 |

Table 3: Two-way ANOVA for analyzing the effects of physiological noise on functional connectivity on the dataset where ROIs were selected based on ATLAS.

| Effects | DF | Sum Sq | Mean Sq | F value | Pr>F | Significancy |
|---|---|---|---|---|---|---|
| Pre-Processing | 1 | 0.0830 | 0.082971 | 3.4150 | 0.0665 | . |
| Area | 5 | 1.1121 | 0.222412 | 9.1543 | 1.177e-07 | *** |
| Pre-Processing:Area | 5 | 0.0338 | 0.006759 | 0.2782 | 0.9245 | |
| Residuals | 156 | 3.7902 | 0.024296 | | | |
| | | | | | | |
| Signif. codes | 0 "***" | 0.001 "**" | 0.01 "*" | 0.05 "." | 0.01 " " | 1 |

Table 4: Two-way ANOVA for analyzing the effects of physiological noise on functional connectivity on the dataset where ROIs were selected based on gICA results.

| Effects | DF | Sum Sq | Mean Sq | F value | Pr>F | Significancy |
|---|---|---|---|---|---|---|
| Pre-Processing | 1 | 0.04257 | 0.042574 | 2.4743 | 0.1177 | |
| Area | 5 | 1.20912 | 0.241823 | 14.0545 | 2.419e-11 | *** |
| Pre-Processing:Area | 5 | 0.01292 | 0.002585 | 0.1502 | 0.9797 | |
| Residuals | 156 | 2.68415 | 0.017206 | | | |
| | | | | | | |
| Signif. codes | 0 "***" | 0.001 "**" | 0.01 "*" | 0.05 "." | 0.01 " " | 1 |

Table 5: Two-way ANOVA for analyzing the effects of physiological noise on functional connectivity on the dataset where ROIs were selected based on the Ind-ICA approach.

| Effects | DF | Sum Sq | Mean Sq | F value | Pr>F | Significancy |
|---|---|---|---|---|---|---|
| ROI-Strategy | 2 | 0.1374 | 0.068710 | 3.4720 | 0.03266 | * |
| Area | 5 | 1.4661 | 0.293216 | 14.8166 | 1.255e-12 | *** |
| ROI-Strategy:Area | 10 | 0.0603 | 0.006029 | 0.3047 | 0.97952 | |
| Residuals | 234 | 4.6308 | 0.019790 | | | |
| | | | | | | |
| Signif. codes | 0 "***" | 0.001 "**" | 0.01 "*" | 0.05 "." | 0.01 " " | 1 |

Table 6: Two-way ANOVA for analyzing the effects of ROI criteria on functional connectivity on the non-physiologically corrected dataset.

| Effects | DF | Sum Sq | Mean Sq | F value | Pr>F | Significancy |
|---|---|---|---|---|---|---|
| ROI-Strategy | 2 | 0.1543 | 0.07717 | 3.6664 | 0.02705 | * |
| Area | 5 | 1.9650 | 0.39301 | 18.6723 | 1.324e-15 | *** |
| ROI-Strategy:Area | 10 | 0.0606 | 0.00606 | 0.2880 | 0.98344 | |
| Residuals | 234 | 4.9251 | 0.02105 | | | |
| | | | | | | |
| Signif. codes | 0 "***" | 0.001 "**" | 0.01 "*" | 0.05 "." | 0.01 " " | 1 |

Table 7: Two-way ANOVA for analyzing the effects of ROI criteria on functional connectivity on the physiologically corrected dataset.

# DISCUSSION AND CONCLUSIONS

In the present study, we investigate how resting state fMRI connectivity estimates within key nodes of the default mode network is affected by two factors: criteria for defining the ROI for the nodes and physiological noise correction.

Regarding ROIs selection strategies, we found a significant effect of ROI criteria on the pairwise correlation between ROIs within DMN. This means that a small change in anatomical location of a region induced by ROI criteria can have an effect on resting-state functional connectivity within the DMN. As expected, ROIs selection based on single-subject DMN nodes estimation provided higher functional connectivity between nodes. Our results are similar to those reported by a recent study (1.5T, 3.4 x 3.4 mm$^2$ voxels, 10 min resting state, TR=2s) from Marrelec & Fransson (2011), who found that relative to atlas and group ICA coordinates, DMN nodes from single subject ICA gave significantly higher functional connectivity estimates. Overall the results are consistent with the expectation that a higher degree of functional gray matter specificity in the network node selection will discard unrelated voxels which introduce uncorrelated fluctuations thereby reducing the overall mean functional connectivity between that node and other nodes.

With regards to the effects of physiological noise correction, we found that there were systematic and significant reductions of functional connectivity amongst all pairs of nodes within the DMN. BOLD signal changes induced by cardiac and respiration processes can lead to false positive or false negative correlations that are not neuronal. Cardiac pulsation results in brain tissue movement and inflow effects increase the correlation of signal fluctuations between areas located near large blood vessels.

Respiration leads to the modulation of the magnetic field that distorts the acquired MR images. Changes in breathing depth and rate can also cause $CO_2$ fluctuations, which are correlated with the BOLD fMRI signal fluctuations (Birn et al., 2006, 2009; Birn, 2012; Buckner et al., 2013; Chang et al., 2009; Dagli et al., 1999). Correction of these physiological artifacts, as Chang et al. has shown, will decrease the magnitude of correlations between precuneus/PCC and other nodes of DMN connectivity (3T, 3.4mm x 3.4mm x 4 mm voxels, 8 min resting state, TR=2s) (Chang & Glover, 2010). Different to our results, they did not find significant effects of functional connectivity reduction between nodes that excluded PCC. However, it is known that the contributions of physiological noise to the total time series is greater in higher fields (Hutton et al., 2011). Therefore our 4T results are consistent with the general finding that it's very important to regress out these physiological artifacts in resting state functional connectivity analysis, especially when working at high fields.

As regards to ROI size, one might think that ROI with a radius of 12 mm is so liberal especially for ATLAS based ROI but considering the result of gICA ROIs selection, we see that pairwise correlation between some ROIs are very close to each other in these two approaches, particularly between MPFC-PCC, PCC-RLPC and rLPC-lLPC (Figure 8). Anyway a detailed analysis of the effect of ROI size on the functional connectivity is beyond the scope of the present study. So, a follow up study can be the potential effect of ROI size on the functional connectivity measures.

To summarize and conclude, this study used 15 young healthy subjects scanned at 4T to estimate resting state functional connectivity within the default node network. The data were available from a different project. The goals of this study were to evaluate the significance of two pre-processing steps, the criteria to select the default mode network nodes and the use of physiological noise correction. We found that both effects were significant. The use of subject-specific ROIs defined from ICA leads to higher connectivity estimates between all nodes relative to the use of atlas-based or group-ICA-based ROIs. The use of physiological noise correction was found to significantly reduce the functional connectivity estimates between all DMN nodes. These results contribute to stress the importance of the choice of in the pre-processing steps in functional connectivity analysis.

## ACKNOWLEDGEMENTS

The authors would like to acknowledge Nicola De Pisapia for sharing with us the data used for this Thesis.